\documentstyle[11pt,newpasp,twoside,epsf]{article}
\def\edcomment#1{\iffalse\marginpar{\raggedright\sl#1\/}\else\relax\fi}
\reversemarginpar

\begin{document}
\title{New Pulsars from Arecibo Drift Scan Searches}
\author{M.~A.~McLaughlin$^1$, D.~R.~Lorimer$^1$, Z.~Arzoumanian$^2$, D.~C.~Backer$^3$, J.~M.~Cordes$^4$,
A.~Fruchter$^5$,
A.~N.~Lommen$^6$ \& K.~Xilouris$^7$} 
\affil{$^1$U. of Manchester, Jodrell Bank Observatory, Cheshire,
SK11~9DL, UK}
\affil{$^2$Laboratory for
        High-Energy Astrophysics, NASA-GSFC, Code 662, Greenbelt, MD
        20771. NAS/NRC Research Associate}
\affil{$^3$Astronomy Department, University of California, Berkeley, CA 94720}
\affil{$^4$Astronomy Department, Cornell University, Ithaca, NY 14853}
\affil{$^5$Space Telescope Science Institute, 3700 San Martin Drive, Baltimore, MD 21218}
\affil{$^6$Anton Pannekoek Instituut, University of Amsterdam}
\affil{$^7$University of Virginia, Department of Astronomy, PO Box 3818, Charlottesville, VA 22903-0818, USA}

\begin{abstract}
We report new pulsars discovered in drift-scan data taken by two collaborations (Berkeley/Cornell
and STScI/NAIC) during the latter stages of the Arecibo upgrade period. The data were taken
with the Penn State Pulsar Machine and are being processed on the COBRA cluster
at Jodrell Bank. Processing is roughly 70\% complete and has resulted in the detection of
10 new and 31 known pulsars, in addition to a number of pulsar candidates. The 10 new pulsars
include one pulsar with a spin-period of 55 ms and another with a spin period of 5.8 ms. At
the completion of the processing, we expect to have discovered roughly 20 new pulsars. All new
pulsars are being subjected to a program of followup observations at Arecibo
to determine spin and 
astrometric parameters.  
\end{abstract}

\section{The Data}

The data discussed were taken during the later stages of the Arecibo upgrade
in 1997 and 1998. These data were taken with the 430-MHz line feed in drift-scan mode,
where sources transit the 430-MHz beam in roughly 42 seconds. Moving the telescope once a day
allowed successive declination strips to be surveyed. Drift-scan surveys such as this have been very  
successful at discovering millisecond pulsars, with both the `planet pulsar' B1257+12 (Wolszczan 
1992) and the
relativistic binary pulsar B1534+12 (Wolszczan 1991) discovered in this mode.
 The data discussed here were taken with the
Penn State Pulsar Machine (PSPM) with 8 MHz of bandwidth,
128 frequency channels and 80~$\mu$s sampling. Our 
8-sigma sensitivity is approximately 1 mJy for low-DM,
long-period pulsars. In total, 130,000 overlapping beams (or roughly 2700 deg$^2$) of data were
collected by our collaboration in this mode. 

\section{Processing}

These data are being processed on the 182-node COBRA cluster at Jodrell Bank.
 This first step of the search processing is
dedispersion over a range of trial dispersion measures. We search to DMs twice those predicted by the
Taylor \& Cordes (1993) model for Galactic electron density, to allow for any uncertainties in the model
or unmodeled clumps of higher electron density. We mitigate the effects of RFI by inspecting the FFTs of
the 0-DM time series of all the beams on a tape (typically 126 beams). All events with
intensities greater than 6 sigma are recorded and any spectral bins which occur 6 times or more are
excised.
To search for periodicities,
an FFT of each dedispersed time series is taken, with summing of up to 16 harmonics. The
threshold signal-to-noise for the periodicity search is 8. 
We also search all dedispersed time series for isolated pulses above a signal-to-noise threshold of 4.
Each dedispersed time series is smoothed successive times to increase sensitivity to
broadened pulses.

\section{Current Status and Results}

Roughly 70\% of all of the data, or roughly 1900 deg$^{2}$, has been processed thus far.
In Figure~1, we show the sky coverage
of the processed data.

\begin{figure}[h]
\plotone{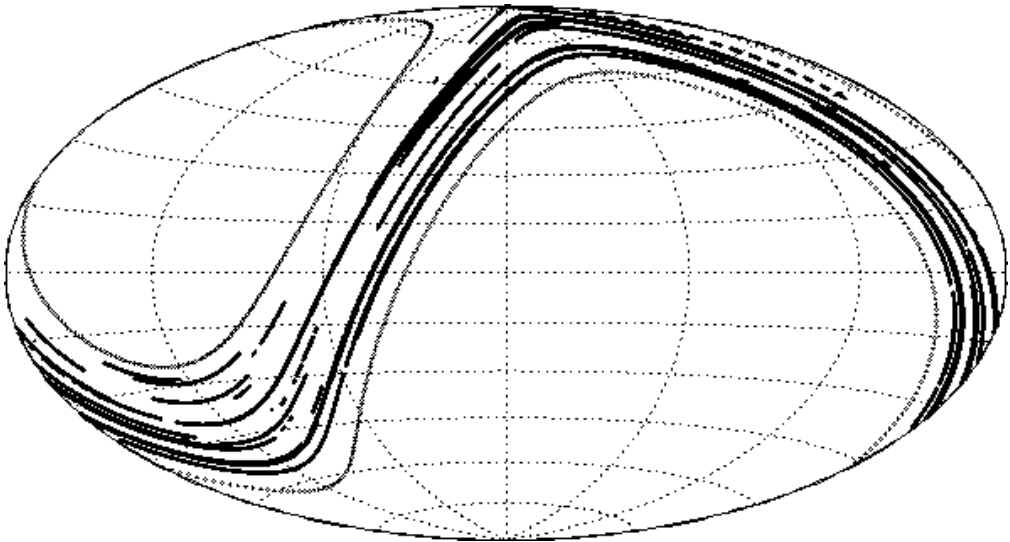}
\caption{Sky coverage of processed drift-scan data. Our data cover roughly 16\% of the entire
Arecibo sky.}
\end{figure}

In the data processed thus far, we have re-detected 31 known pulsars and discovered 10 new pulsars.
Four of the 31 known pulsars and one of the 10  new pulsars are millisecond pulsars.
Roughly half of all pulsars were detected in both the periodicity and single-pulse searches. Two
known pulsars, J1918+1444 (P = 1.18 s) and J1915+0738 (P = 1.54 s) were detected as strong sources
 in the single-pulse search
but were not detected in the periodicity search. This illustrates the importance of performing
both types
of search analyses, especially for pulsars with long periods and for short integration times. 

Including the data reported here, drift-scan searches done with Arecibo over the upgrade
 period (Foster et al. 1995, Camilo et al. 1996,
 Ray et al. 1996, Lommen et al. 2000)
 have covered roughly 50\% of the
entire Arecibo sky and have resulted in the discovery of 139 pulsars. Detection rates of
of one pulsar per 36 $-$ 47 deg$^{2}$ and one millisecond pulsar per 250 $-$ 700 deg$^{2}$
have been achieved in these searches.

In Table 1, we list the new pulsars discovered in our data along with their periods, dispersion measures and detected
signal-to-noises. Timing observations have not yet been scheduled for these pulsars, so
period derivatives are not available for most of them. Preliminary timing data for
PSR~J0610+21 indicate that it is an isolated pulsar with a period derivative of $5.3\times10^{-16}$
and inferred age
and magnetic field of 2~Myr and $2\times10^{11}$~Gauss, respectively. Timing residuals are shown in
Figure~2. This pulsar is in a fairly
unique place in the $P-\dot{P}$ diagram and has parameters similar to PSR~B1259$-$63, the
pulsar-Be star binary system. Perhaps PSR~J0610+21 experienced a similar evolutionary history
but somehow lost its companion.

\begin{figure}[h!b]
\plotone {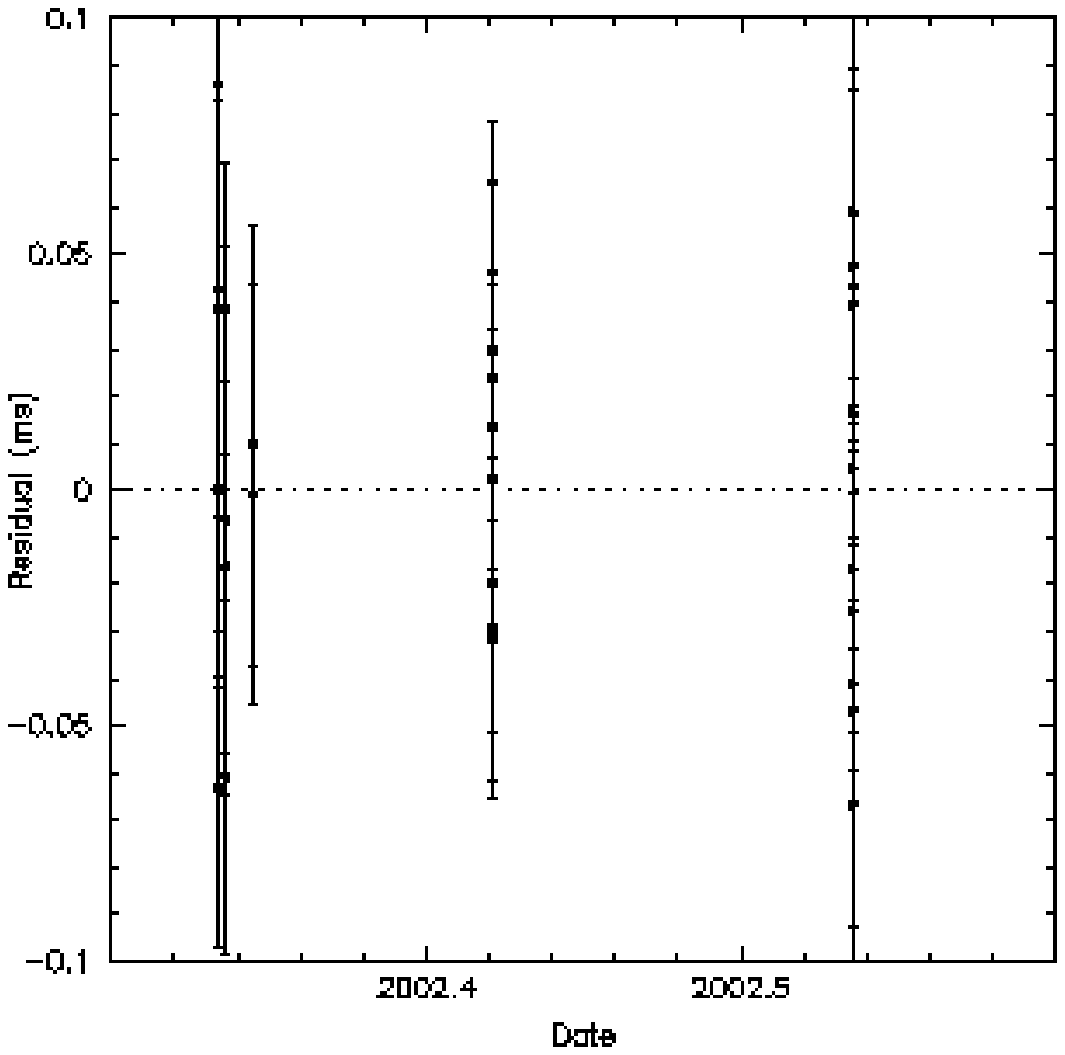}
\caption{Timing residuals for PSR~J0610+21 for a model consisting of $P$, $\dot{P}$, RA and DEC. Timing
data have been collected for seven epochs.}
\end{figure}

\begin{center}
\begin{tabular}{|lccc|} 
\multicolumn{4}{c}{Table 1: New Pulsar Parameters} \\
\hline
Name & P (ms) & DM (pc cm$^{-3}$) & SNR \\
\hline
J0152+09 &	915	& 20	& 9	\\
J0610+21 &	55.6	& 40	& 9	\\
J0815+09 &	645	& 20	& 19	\\
J0832+17 &	865	& 38	& 39	\\
J1437+07 &	395	& 37	& 11	\\
J1453+19 &	5.79	& 14    & 27	\\
J1504+21 &	2205	& 13	& 14	\\
J1746+22 &	3459	& 100	& 10	\\
J1823+06 &	752	& 64	& 14	\\
J2045+09 &	395	& 32	& 36	\\
\hline
\end{tabular}
\end{center}

\acknowledgements
We thank operators at Arecibo for their help collecting this data.
 MAM is an NSF MPS-DRF Postdoctoral Fellow. DRL is a University Research Fellow funded by
the Royal Society.

\end{document}